\documentclass[showpacs,twocolumn]{revtex4}

\usepackage{graphicx}
\usepackage{dcolumn}
\usepackage{amsmath}

\begin{document}
\title{Accretion onto a higher dimensional black hole}

\author{Anslyn J. John$^a$}\email{johnan@ukzn.ac.za}
\author{Sushant G. Ghosh$^{a,\;b}$}\email{sghosh2@jmi.ac.in}
\author{Sunil D. Maharaj$^{a}$}\email{maharaj@ukzn.ac.za}

\affiliation{$^a$Astrophysics and Cosmology Research Unit,
 School of Mathematics, Statistics and Computer Science,
 University of KwaZulu-Natal, Private Bag X54001,
 Durban 4000, South Africa}

\affiliation{$^b$Center for Theoretical Physics,
 Jamia Millia Islamia,  New Delhi 110025
 India}

\begin{abstract}
We  examine  the steady-state spherically symmetric  accretion of relativistic fluids, with a polytropic equation of state, onto a higher dimensional Schwarzschild black hole. The mass accretion rate, critical radius, and flow parameters are determined and compared with results obtained in standard four dimensions.   The accretion rate, $\dot{M}$, is an explicit function of the black hole mass, $M$, as well as the gas boundary conditions and the dimensionality, $D$, of the spacetime. We also find the asymptotic compression ratios and temperature profiles below the accretion radius and at the event horizon. This analysis is a generalization of Michel's solution to higher dimensions and of the Newtonian expressions of Giddings and Mangano which considers the accretion of TeV black holes.
\end{abstract}

\maketitle

\section{Introduction}
Accretion of matter onto black holes is an extensively studied topic and the most likely scenario to explain the high energy output from active galactic nuclei and quasars.  The seminal paper by Bondi \cite{bondi} is devoted to formulating the theory of stationary, spherically symmetric and transonic accretion of adiabatic fluids onto astrophysical objects.  Indeed  Bondi \cite{bondi} solved the problem of a polytropic gas accreting onto a central object under the influence of gravity, which generalizes the earlier results of Bondi and Hoyle  \cite{bhoyle} and Hoyle and Lyttleton \cite{lyttleton} who investigated  pressure-free gas being dragged onto a massive central object. There has been some confusion in distinguishing these cases in the literature but the latter case is usually referred to as Lyttleton-Hoyle accretion whilst the former is termed Bondi accretion \cite{edgar}. The key distinction between the two cases is that the gas and the accretor are in the same inertial rest-frame in Bondi accretion whilst in Lyttleton-Hoyle accretion the gas has a finite velocity at infinity \cite{edgar}. Both of these cases are formulated in the context of Newtonian gravity. There exists a large body of literature devoted to theoretical and observational studies of accretion processes. Detailed treatments can be found in any of the standard texts  \cite{frank}, \cite{st}, or \cite{shu}.

In the framework of general relativity, the steady-state spherically symmetric flow of a test gas onto a Schwarzschild black hole was investigated by Michel \cite{michel}. Following Michel's relativistic generalization of Bondi's treatment, Begelman \cite{begelman} discussed some aspects of the critical points of the accretion problem. Spherical accretion and winds in general relativity have also been considered using equations of state other than the polytropic one. Other comprehensive treatments include the calculation of the luminosity and frequency spectrum \cite{shap73a}, the influence of an interstellar magnetic field on the accretion of ionized gases \cite{shap73b}, and accreting processes onto a rotating black hole \cite{shap74}. Various radiative processes have been incorporated by  Blumenthal  and Mathews \cite{blum}, and Brinkmann \cite{brink}. Recently Malec \cite{malec}  provided the solution for general relativistic spherical accretion with and without back-reaction and showed that relativistic effects enhance mass accretion when back-reaction is neglected.  Accretion onto a charged black hole was considered in \cite{michel} and more fully investigated in \cite{charge}. Spherical winds and shock transitions were studied in \cite{blum}.  Detailed studies of spherically symmetric accretion of different types of fluids onto black holes were further undertaken in a number of works.

String theory is widely believed to be the most promising candidate for a unified description of everything; it predicts the existence of higher dimensions for its consistency. It is important to note that gravity in higher dimensions is very different from its four dimensional counterpart. For example, black holes with a fixed mass may have arbitrarily large angular momentum \cite{mp}.  There is a growing
realization  that the physics of higher dimensional black holes can be markedly
different, and much richer than in four dimensions \cite{empran,horwitz}. The higher dimensional theories are applied in various unsolved problems in astrophysics and cosmology to gain new insights into them.   On the other hand, the accretion process is a powerful way of understanding  the physical nature of the central celestial condensed objects; the analysis of signatures of the  accretion disk could provide another possible way to detect the influence of higher dimensions.

An interesting study of accretion in higher dimensions  was initiated by Giddings and Mangano, \cite{tev}, who studied higher dimensional accretion onto TeV black holes in the Newtonian limit. They argued that since the Bondi radius is typically much larger than the event horizon radius one can safely describe the gravitational field of the accreting black hole by a Newtonian potential. In this paper we look at steady, spherical accretion of a polytropic gas onto a point mass in $D$-dimensional general relativity. The event horizon radius is of the order ${GM}/{c^2}$ whilst the Bondi radius scales like ${GM}/{a^{2}}$ where $a$ is the fluid sound speed. If the sound speed approaches the speed of light i.e. $a \sim c$ then the Bondi radius approaches the event horizon and one can no longer disregard the effects of spacetime curvature. Such a situation could arise if, for example, the accreting fluid obeyed a stiff equation of state. Higher dimensional black holes arise in extensions to the standard model of particle physics like $M$-theory. We consider $D$-dimensional general relativity rather than $D$-dimensional Newtonian gravity as a low energy limit of semi-classical physics.

The accretion of phantom matter onto $5$-dimensional charged black holes was investigated by Sharif and Abbas \cite{phantom}.
As was the case in phantom accretion onto
$4$-dimensional Schwarzschild \cite{babprl} and Reissner-Nordstrom black holes \cite{phantomrn}, the black hole mass decreases.
The accretion of phantom energy onto Einstein-Maxwell-Gauss-Bonnet black holes was studied in \cite{jamhus}. They showed that the evolution of the black hole mass was independent of its mass and depends only on the energy density and pressure of the phantom energy. This result is similar to that arising in the accretion of phantom matter onto a 2+1 dimensional Banados-Teitelboim-Zanelli (BTZ) black hole \cite{btz}. Jamil and Hussain \cite{jamhus} conjecture that the rate of change of mass of a black hole due to phantom energy accretion depends on the mass of a black hole in 4 dimensions only.

We believe that this mass independence is not a generic feature of higher dimensional accretion and arises because of the pathological nature of phantom matter which violates the dominant energy condition. We consequently focus our attention on more conventional fluids. In this paper we formulate and solve the problem of a polytropic gas accreting onto a Schwarzschild black hole in arbitrary dimensions. The gravitational model used is $D$-dimensional general relativity. We determine analytically the critical radius, critical fluid velocity and sound speed, and subsequently the mass accretion rate. We then obtain expressions for the asymptopic behaviour of the fluid density and temperature near the event horizon.

We use the following values for physical constants and the accreting system's parameters:
$c = 3.00 \times 10^{10} \mathsf{cm. s}^{-1}$,
$G = 6.674 \times 10^{-8} \mathsf{cm}^{3}. \mathsf{g.}^{-1} \mathsf{s}^{-2}$,
$k_B = 1.380 \times 10^{-16} \mathsf{erg. K}^{-1}$,
$M = M_{\odot} = 1.989 \times 10^{33} \mathsf{g}$,
$m = m_{p} =1.67 \times 10^{-24} \mathsf{g}$,
$a_{\infty} = 10^6 \mathsf{cm.s}^{-1}$,
$n_{\infty} = 1 \mathsf{cm}^{-3}$,
$T_{\infty} = 10^2 \mathsf{K}$.

\section{Basic equations}
In this model, we assume a higher dimensional spherically symmetric metric, and steady state flow onto a nonrotating black hole of mass $M$ at rest in the interstellar medium.  A black hole surrounded by the stellar plasma is expected to capture matter within a certain distance, which may result in a flow of plasma towards the black hole.  It may be recalled that the non-relativistic model was discussed by Bondi \cite{bondi} and the  standard 4-dimensional general relativistic version was developed by Michel \cite{michel}.  The known analytic relativistic accretion solution onto the Schwarzschild black hole by Michel \cite{michel} is extended to arbitary dimensions.  To  do this, we assume that the black hole is represented by  the $D$-dimensional generalization of Schwarzschild black hole described by the line element \cite{tangherlini1963schwarzschild}
 \begin{eqnarray}
ds^2 &=& - \left(1- \frac{\mu}{r^{D-3}} \right) dt^2 + \left(1- \frac{\mu}{r^{D-3}} \right)^{-1} dr^2  \nonumber\\
&& + r^2 d \Omega^{2}_{D-2}, \label{metricHD}
\end{eqnarray}
where $D$ is the spacetime dimension and $\mu$ is related to the mass of the black hole via $\mu = \frac{16\pi GM}{(D-2) \Omega_{D-2}}$. Here $\Omega_{D-2}$ is the area of a unit $(D-2)$-dimensional sphere
\[
 \Omega_{D-2} =   \frac{2 \pi^{\frac{D-1}{2}}}{\Gamma (\frac{D-1}{2})},
\]
 and $d\Omega^{2}_{D-2}$ is the line element on the sphere, viz.
 \[
d \Omega^{2}_{D-2} = d \theta^{2}_{1} + \sum^{D-2}_{n=2}  \left(  \prod^{n}_{m=2} \sin^2 \theta_{(m-1)} \right) d \theta^{2}_{n}.
\]
We use comoving coordinates $ (x^\alpha) = (t, r, \theta_1, \theta_2, \cdots , \theta_{D-2} )$
and units where $c=G=1$. At particular points we
reintroduce $c$ and $G$ to ease comparison with
Shapiro and Teukolsky \cite{st}. The above metric (\ref{metricHD}) provides a vacuum solution in the $ D $-dimensional Einstein theory, describing static, asymptotically flat higher dimensional black holes. As seen easily, the black hole horizon exists at $r = r_H \equiv \mu^\frac{1}{D-3}$. The standard Schwarzschild solution can be recovered by setting $D=4.$

Next,  we probe how the extra dimensions  affect the Bondi accretion rate, the asymptotic compression ratio, and the temperature profiles. We consider the steady-state radial inflow of gas onto a central mass $M$. The gas is approximated as a perfect fluid described by the energy momentum tensor \begin{equation}
T^{\alpha\beta} = \left( \rho + p \right)u^\alpha u^\beta+ p g^{\alpha\beta} \label{stress}
\end{equation}
where  $\rho$ and $p$ are the fluid proper energy density and pressure respectively, and
\begin{equation}
u^\alpha = \frac{d x^\alpha }{ds}, \label{fluid}
\end{equation}
 is the fluid $D$-velocity, which obeys the normalization condition $u^\alpha u_ \alpha = -1.$  We also define the proper baryon number  density  $n$, and the  baryon number flux $J^\alpha  = n u^\alpha$. All these quantities are measured in the local inertial rest frame of the fluid. The spacetime curvature is dominated by the compact object and we ignore the self-gravity of the fluid. If no particles are created or destroyed then particle number is conserved and
  \begin{equation}
\nabla_\alpha J^\alpha = \nabla_\alpha ( n u^\alpha ) = 0, \label{masscon}
\end{equation}
where $\nabla_\alpha$ denotes the covariant derivative with respect to the coordinate $x^\alpha$. Conservation of energy and momentum is governed by
 \begin{equation}
\nabla_\alpha T^{\alpha}_{\beta} = 0. \label{momentum}
\end{equation}
We define the radial component of the $D$-velocity, $v(r) = u^1 = dr/ds $. Since $u_\alpha u^\alpha = -1$ and the velocity components vanish for $\alpha > 1$ we have \begin{equation}
(u^0)^2 = \frac{v^2 + 1 - \frac{\mu}{r^{D-3}}}{\left( 1 - \frac{\mu}{r^{D-3}}\right)^2}. \label{unought}
\end{equation}
Equation (\ref{masscon}) for, our $D$-dimensional Schwarzschild case,  can be written as \begin{equation}
\frac{1}{r^{D-2}} \frac{d}{dr} \left( r^{D-2} n v \right)= 0. \label{masscon1}
\end{equation}
Our assumption of spherical  symmetry and steady-state flow makes (\ref{momentum}) comparatively easier to tackle. The $\beta = 0$ component of (\ref{momentum}) is \begin{equation}
\frac{1}{r^{D-2}} \frac{d}{dr} \left( r^{D-2} (\rho + p) v \left(  1 - \frac{\mu}{r^{D-3}} +   v^2 \right)^{1/2} \right) = 0. \label{momentumzero}
\end{equation}
The $\beta=1$ component can be simplified to
\begin{equation}
v \frac{d v }{dr} = - \frac{d p}{d r} \left( \frac{ 1 - \frac{\mu}{r^{D-3}} + v^2  }{\rho + p} \right) - \frac{D-3}{2}\frac{\mu}{r^{D-2}}. \label{momentumone}
\end{equation}
These expressions generalise, to arbitrary dimensions $D$, those obtained by Michel \cite{michel} for spherical accretion onto a Schwarzschild black hole and we recover them in the 4-dimensional limit:
\begin{eqnarray}
\frac{1}{r^{2}} \frac{d}{dr} \left( r^{2} n v \right) &=& 0, \label{GR-old-a}\\
\frac{1}{r^{2}} \frac{d}{dr} \left( r^{2} (\rho + p) v \left( 1- \frac{2M}{r} + v^2 \right)^{1/2} \right) &=& 0, \label{GR-old-b} \\
v \frac{d v }{dr} = - \frac{d p}{d r} \left( \frac{1 -\frac{2M}{r} + v^2} {\rho + p} \right) &-& \frac{M}{r^{2}}. \label{GR-old-c}
\end{eqnarray}

\begin{table}

\caption{\label{gamma1p1} Critical radius $r_s$, accretion parameter $\lambda_s$ and accretion rate $\dot{M}$ for $\gamma =1.1$. Values are expressed in geometrized units.}
\begin{tabular}{|c|c|c|c|}
\hline $D$ & $r_s$ & $ \lambda $  & $\dot{M}$  \\
\hline 4 & 8.453$ \times 10^{20} $ & 0.2488 & 8.262$ \times 10^{25} $ \\
\hline 5 & 2.250$ \times 10^{10} $ & 0.4967 & 1.136$ \times 10^{15} $ \\
\hline 6 & 6.886$ \times 10^6  $& 0.6004 & 2.463$ \times 10^{11} $ \\
\hline 7 & 1.243$ \times 10^5 $ & 0.6543 & 3.482$ \times 10^9  $\\
\hline 8 & 1.143$ \times 10^4 $ & 0.6868 & 2.641$ \times 10^8 $ \\
\hline 9 & 2.367$ \times 10^3 $  & 0.7085 & 4.660$ \times 10^7 $ \\
\hline 10 & 7.779$ \times 10^2 $ & 0.7239 & 1.335$ \times 10^7 $ \\
\hline 11 & 3.407$ \times 10^2 $ & 0.7353 & 5.186$ \times 10^6 $ \\
\hline
\end{tabular}

\end{table}

\begin{table}

\caption{\label{gamma4p3} Critical radius $r_s$, accretion parameter $\lambda_s$ and accretion rate $\dot{M}$ for $\gamma =1.1$ for $\gamma =4/3$. Values are expressed in geometrized units.}
\begin{tabular}{|c|c|c|c|}
\hline $D$ & $r_s$ & $ \lambda $  & $\dot{M}$  \\ 
\hline 4 & 4.972$ \times 10^{20} $ & 0.1768 & 5.871$ \times 10^{25} $ \\
\hline 5 & 1.937$ \times 10^{10} $ & 0.4330 & 9.902$ \times 10^{14} $ \\
\hline 6 & 6.360$ \times 10^6  $& 0.5473 & 2.246$ \times 10^{11} $ \\
\hline 7 & 1.179$ \times 10^5 $ & 0.6077 & 3.234$ \times 10^9 $ \\
\hline 8 & 1.100$ \times 10^4 $  & 0.6444 & 2.478$ \times 10^8 $ \\
\hline 9 & 2.296$ \times 10^3 $ & 0.6689 & 4.400$ \times 10^7  $\\
\hline 10 & 7.587$\times 10^2 $ & 0.6864 & 1.266$ \times 10^7  $\\
\hline 11 & 3.336$\times 10^2 $ & 0.6994 & 4.932$ \times 10^6  $\\
 \hline
\end{tabular}
\end{table}

\begin{table}

\caption{\label{gamma1p6} Critical radius $r_s$, accretion parameter $\lambda_s$ and accretion rate $\dot{M}$ for $\gamma =1.1$ for $\gamma =1.6$. Values are expressed in geometrized units.}
\begin{tabular}{|c|c|c|c|}
\hline $D$ & $r_s$ & $ \lambda $  & $\dot{M}$  \\
\hline 4 & 9.945$ \times 10^{19} $ & 0.09174 & 3.047$ \times 10^{25} $ \\
\hline 5 & 1.500$ \times 10^{10} $ & 0.3545 & 8.107$ \times 10^{14} $ \\
\hline  6 & 5.626$ \times 10^6 $ & 0.4818 & 1.977$ \times 10^{11}  $\\
\hline  7 & 1.091$ \times 10^5 $ & 0.5503 & 2.928$ \times 10^9 $ \\
\hline  8 & 1.040$ \times 10^4 $ & 0.5922 & 2.277$ \times 10^8 $ \\
\hline  9 & 2.199$ \times 10^3 $ & 0.6202 & 4.080$ \times 10^7  $\\
\hline  10 & 7.325$ \times 10^2 $ & 0.6403 & 1.181$ \times 10^7 $ \\
\hline  11 & 3.238$ \times 10^2 $ & 0.6553 & 4.620$ \times 10^6  $\\
\hline
\end{tabular}
\end{table}

\section{Analysis}
In the spirit of the original calculation of Bondi \cite{bondi} we obtain the mass accretion rate from a qualitative analysis of (\ref{masscon1}) and (\ref{momentumone}). For an adiabatic fluid there is no entropy production and the conservation of mass-energy is governed by \begin{equation}
T ds = 0 = d \left( \frac{\rho}{n}\right) + p\; d \left(\frac{1}{n}\right), \label{energy}
\end{equation}
which implies the relation \begin{equation}
\frac{d \rho}{d n} = \frac{\rho + p}{n}. \label{energycon}
\end{equation}
As in the 4-dimensional case, we define the adiabatic sound speed $a$ \cite{st},  via  \begin{eqnarray}
a^2 &\equiv& \frac{d p}{d \rho}
= \frac{d p}{d n} \frac{n}{\rho + p}. \label{asqrd}
\end{eqnarray}
where we have used equation (\ref{energycon}).
The continuity and momentum equations, for the $D$-dimensional Schwarzschild case, may be written as
 \begin{eqnarray}
\frac{1}{v}v' + \frac{1}{n}n' &=& -\frac{D-2}{r}, \label{numbden} \\
v v' + \left( 1 - \frac{\mu}{r^{D-3}} + v^2\right) \frac{a^2}{n} n' &=& - \frac{D-3}{2} \frac{\mu}{r^{D-2}}, \label{vel}
\end{eqnarray}
where  a prime ($'$) denotes a derivative with respect to $r$, and we obtain this system
  \begin{eqnarray}
v' &=& \frac{N_1}{N}, \nonumber \\
n' &=& -\frac{N_2}{N}, \label{firstorder}
\end{eqnarray}
where
\begin{eqnarray} \label{d1}
N_1 &=& \frac{1}{n} \left( \left(1 - \frac{\mu}{r^{D-3}} + v^2 \right) (D-2)\frac{a^2}{r}  \right. \nonumber \\
 && \left.  - \frac{D-3}{2} \frac{\mu}{r^{D-2}}  \right), \label{d1-a} \\
N_2 &=& \frac{1}{v} \left( (D-2) \frac{v^2}{r} - \frac{D-3}{2} \frac{\mu}{r^{D-2}}  \right), \label{d1-b} \\
N &=& \frac{v^2 - \left( 1 - \frac{\mu}{r^{D-3}} + v^2 \right)a^2 }{vn}. \label{d1-c}
\end{eqnarray}
At large $r$ we demand the flow be subsonic i.e. $v < a$ and
since the sound speed is always subluminal i.e. $a < 1$,  we have $v^2 \ll 1$. The denominator (\ref{d1-c}) is
thus \begin{equation}
N \approx \frac{v^2 - a^2}{v n}
\end{equation}
and so $N > 0$ as $r \rightarrow \infty $.
At the event horizon $r_H = \mu^\frac{1}{D-3}$, we have \begin{equation}
N = \frac{v^2 (1-a^2)}{v n}.
\end{equation}
Under the causality constraint $a^2<1$,  we have $N <0$, therefore we must have $N=0$ for some critical point
$r_s$ where $r_H < r_s < \infty$. Interestingly, the expression for $N$ is exactly the same  in 4-dimensions \cite{st}. The flow must pass through a critical point outside the event horizon. In order to avoid discontinuities in the flow, we must have $N=N_1=N_2=0$ at $r=r_s$, i.e.

\begin{eqnarray}\label{n1}
N_1 &=& \frac{1}{n_s} \left( \left(1 - \frac{\mu}{r_s^{D-3}} + v^{2}_{s} \right) \frac{D-2}{r_s} a^{2}_{s}  \right. \nonumber \\
& & \left. - \frac{D-3}{2} \frac{\mu}{r^{D-2}_{s}}  \right) = 0, \label{n1-a} \\
N_2 &=& \frac{1}{v_s} \left( (D-2) \frac{v^{2}_{s}}{r_s} - \frac{D-3}{2} \frac{\mu}{r^{D-2}_{s}}  \right) = 0, \label{n1-b} \\
N &=& \frac{v^{2}_{s} - \left( 1 - \frac{\mu}{r_s^{D-3}} + v^{2}_{s} \right)a^{2}_{s} }{v_s n_s} = 0. \label{n1-c}
\end{eqnarray}
where $v_s \equiv v(r_s),\; a_s \equiv a(r_s)$, etc. At the critical point that satisfies equation (\ref{n1}) we have \begin{eqnarray}
v_{s}^{2} &=& \frac{a_{s}^{2}}{1 + \frac{D-1}{D-3} a_{s}^{2}} \label{crit1} \\
&=& \frac12 \frac{D-3}{D-2} \frac{\mu}{r_{s}^{D-3}}. \label{crit2}
\end{eqnarray}
Again in the 4-dimensional limit, we have $v_{s}^{2}=M/r^2$ exactly as in Shapiro and Teukolsky \cite{st}.

We determine the accretion rate, $\dot{M}$, by analyzing the continuity equation at the critical
radius, $r_s$.
 Following Michel \cite{michel}, we write the continuity equation explicitly in the form of a conservation equation.
  Integrating equation (\ref{masscon1}) over a $(D-1)$-dimensional volume and multiplying by $m$, the mass of each gas particle, we obtain

\begin{equation}
\frac{2 \pi^{(D-1)/2}}{\Gamma \left( \frac{D-1}{2}  \right)} r^{D-2} m n v = \dot{M} \label{bondi}
\end{equation}
where $\dot{M}$ is an integration constant, independent of $r$, having dimensions of mass per unit time. $\dot{M}$ is the higher dimensional generalization of Bondi's mass accretion rate.  Note that (\ref{bondi}) reduces to the standard Schwarzschild case when $D=4$,
viz. $\dot{M} = 4 \pi r^2mnv$.  Equations (\ref{masscon1}) and (\ref{momentumzero}) can be combined to yield

\begin{equation}
\left( \frac{\rho + p}{n} \right)^2 \left( 1 - \frac{\mu}{r^{D-3}} + v^2 \right) = \left( \frac{\rho_{\infty} + p_{\infty}}{n_{\infty}}\right)^2, \label{bernoulli}
\end{equation}
which is the $D$-dimensional generalization of the relativistic Bernoulli equation.
Eqs. (\ref{bondi}) and (\ref{bernoulli}) are fundamental conservation equations for the flow of matter onto a $D$-dimensional Schwarzschild black hole where we have ignored the back-reaction of matter. In the 4-dimensional limit these expressions reduce to those obtained by Michel \cite{michel}.

\subsection{The polytropic solution}
In order to calculate explicitly $\dot{M}$, following Bondi \cite{bondi} and Michel \cite{michel}, we introduce the polytrope equation of state \begin{equation}
p = K n^{\gamma}, \label{eos}
\end{equation}
where $K$ is  constant and the adiabatic index satisfies $1 < \gamma < \frac{5}{3}$. On inserting (\ref{eos}) into energy equation (\ref{energy}) and integrating, we obtain
\begin{equation}
\rho = \frac{K}{\gamma -1}n^{\gamma} + m n, \label{energyint}
\end{equation}
where $m$ is an integration constant and $mn$ is the rest--energy density. Using (\ref{asqrd}) we rewrite the Bernoulli equation (\ref{bernoulli}) as
  \begin{eqnarray}
&& \left(1 + \frac{a^{2}}{\gamma - 1 - a^{2} }\right)^2 \left( 1  - \frac{\mu}{r^{D-3}} + v^2 \right) \nonumber \\
&& = \left(1 +  \frac{a_{\infty}^{2}}{\gamma - 1 - a_{\infty}^{2}}\right)^2. \label{bernoulli2}
\end{eqnarray}
At the critical radius $r_s$, this must satisfy \begin{equation}
\left[ \frac{ (D-3) + (D-1)a_{s}^{2}}{(D-3)} \right] \left( 1 - \frac{a_{s}^{2}}{\gamma -1} \right)^2 = \left( 1 - \frac{a_{\infty}^{2}}{\gamma - 1} \right)^2. \label{berncrit}
\end{equation}
where we have used the critical velocity (\ref{crit1}) and sound speed (\ref{crit2}). For large, but finite $r$, i.e. $r \geq r_s$ the baryons should still be non-relativistic, i.e. $T \ll mc^2/k = 10^{13} K$ for neutral hydrogen. In this regime we expect $a \leq a_s \ll 1$. Expanding (\ref{berncrit}) to leading order in $a_s$ and $a_{\infty}$ we obtain \begin{equation}
a_{s}^{2} \approx \frac{2 (D-3)}{(3D -7) -\gamma (D-1) } a_{\infty}^{2}. \label{bernapprox}
\end{equation}
We thus obtain the critical radius $r_s$ in terms of the black hole mass $M$ and the boundary condition $a_{\infty}$:
\begin{eqnarray}
r_{s}^{D-3} &=&  \frac12 \frac{D-3}{D-2} \mu \left[ \frac{1 + \frac{D-1}{D-3} a_{s}^{2} }{a_{s}^{2}} \right]\nonumber \\
&\approx& \frac{1}{4(D-2)} \frac{\mu}{ a_{\infty}^{2}} \left[ (3D -7) - (D-1) \gamma \right] .\label{rcrit}
\end{eqnarray}
Re-introducing the normalised constants this reads,
 \begin{eqnarray}
r_{s}^{D-3} &\approx&  \frac{2}{(D-2)^2} \pi^\frac{3-D}{2}
  \Gamma \left( \frac{D-1}{2}\right)  \nonumber\\
  && \times \left[ (3D-7) -(D-1)\gamma \right]  \frac{G M}{a_{\infty}^{2}} .\label{rcrit1}
\end{eqnarray}
For $D=4$ we recover the familiar Bondi radius for the Schwarzschild solution, viz.
\[
r_s = \frac{5-3\gamma}{4} \frac{GM}{a^2_\infty} .
\]
From (\ref{asqrd}), (\ref{eos}) and (\ref{energyint}) we have \begin{equation}
\gamma K n^{\gamma -1} = \frac{m a^{2}}{1 - a^2/ (\gamma-1)} . \label{nsubs}
\end{equation}
For $a^{2}/(\gamma-1) \ll 1$ we have $n \sim a^{2/(\gamma-1)}$ and \begin{equation}
\frac{n_s}{n_{\infty}} \approx \left( \frac{a_s}{a_{\infty}} \right)^{2/(\gamma-1)}. \label{nsubs2}
\end{equation}
We are now in  a position to evaluate the accretion rate, $\dot{M}$. Since $\dot{M}$ is independent of $r$, equation (\ref{bondi}) must also hold for $r=r_s$. We use the critical point to determine the $D$--dimensional Bondi accretion rate, \begin{eqnarray}
\dot{M} &=& \frac{2 \pi^\frac{D-1}{2}}{\Gamma \left( \frac{D-1}{2}  \right) } r^{D-2} m n v \nonumber \\
&=& \frac{2 \pi^\frac{D-1}{2}}{\Gamma \left( \frac{D-1}{2}  \right) } r^{D-2}_{s} m n_s v_s \nonumber \\
&=& \frac{2 \pi^\frac{D-1}{2}}{\Gamma \left( \frac{D-1}{2}  \right) } \lambda \mu^\frac{D-2}{D-3} m n_{\infty} a_\infty^\frac{1-D}{D-3},  \label{accretion}
\end{eqnarray}
where we have defined the dimensionless accretion eigenvalue
\begin{eqnarray}
\lambda &=&
 \frac{1}{(D-2)^\frac{D-2}{D-3}} \left(\frac{D-3}{2} \right)^\frac{\gamma +1}{2(\gamma -1)}  \nonumber \\
  &&  \times \left[  \frac{ (3D-7) + (1-D)\gamma} {4} \right]^\frac{7-3D+ (D-1)\gamma}{ 2(D-3)(\gamma -1)} . \label{eigenv}
\end{eqnarray}
We rewrite the $D$-dimensional accretion rate explicitly in terms of $G$, the gravitational constant,
\begin{eqnarray}
 \dot{M} &=&
  \sqrt{\pi} \left[ 2^{4D-9} (D-2)^{2(2-D)} \Gamma \left(\frac{D-1}{2} \right) \right]^\frac{1}{D-3} \nonumber \\
&& \times (GM)^\frac{D-2}{D-3}  m n_{\infty} a_\infty^\frac{1-D}{D-3}
  \left(\frac{D-3}{2}\right)^\frac{\gamma +1}{2(\gamma -1)}  \nonumber \\
 && \times \left[ \frac{3D-7 +(1-D)\gamma}{4} \right]^\frac{7-3D+ (D-1)\gamma}{
 2(D-3)(\gamma -1)} . \label{mdotexplicit}
\end{eqnarray}
Note that the accretion rate scales as $\dot{M} \sim M^{(D-2)/(D-3)}$. This extends the familiar result of Bondi \cite{bondi} where $\dot{M} \sim M^2$ and suggests potentially observable hints of the presence of higher dimensions.
The 4-dimensional limit of Eq. (\ref{mdotexplicit}), which reads
\begin{equation}
\dot{M} = 4 \pi   \left( \frac{GM}{a_\infty^2} \right)^2 m n_{\infty} a_{\infty} \left( \frac12 \right)^\frac{\gamma +1}{2(\gamma -1)}
\left( \frac{5-3\gamma}{4}\right)^\frac{3\gamma -5}{2(\gamma - 1)},
\end{equation} 
is exactly the same expression quoted  in Shapiro and Teukolsky \cite{st} when  $1 < \gamma < 5/3.$  For transonic flow the relativistic accretion rate agrees with its Newtonian counterpart to leading order.  The behavior of the critical radius $r_s$, the acccretion parameter $\lambda$, and accretion rate $\dot{M}$, as a function of the dimension $D$, is depicted in Tables I-III for fiducial values of the adiabatic index $\gamma$.  It turns out that the spacetime dimension plays an  important role in the accretion process; interestingly it slows down the accretion rate (see also Figs.~I-II). This may be a useful feature to incorporate in astrophysical applications.

\section{Asymptotic behaviour}
 The results obtained in the previous sections are valid at large distances from the black hole near the critical radius $r_s \gg r_H$.  It turns out that the  general relativistic accretion rate \cite{michel}, to lowest order, is equivalent to the Newtonian transonic flow obtained by Bondi \cite{bondi}. We have proved the results of Michel \cite{michel} carry over to higher dimensions.   Next, we estimate the flow characteristics for $r_H < r \ll r_s$ and at the event horizon $r = r_H$.

\subsection{Sub-Bondi radius}
The gas is supersonic at distances below the Bondi radius so $v>a$ when $r_H < r \ll r_s$. From (\ref{bernoulli2}) we obtain an upper bound on the radial dependence of the gas velocity viz.
 \begin{equation}
v^2 \approx \frac{\mu}{r^{D-3}}. \label{gasvel}
\end{equation}
We now estimate the gas compression on these scales using (\ref{bondi}), (\ref{mdotexplicit}) and (\ref{gasvel}):
 \begin{equation}
\frac{n(r)}{n_{\infty}} =  \lambda \left( \frac{\mu}{a^2_\infty r^{D-3}} \right)^\frac{D-1}{2(D-3)}. \label{gascomp}
\end{equation}
Assuming a Maxwell-Boltzmann gas, $p = n k_B T$, we find the adiabatic temperature profile using (\ref{eos}) and (\ref{gascomp}):
\begin{equation}
\frac{T(r)}{T_{\infty}} = \lambda^{\gamma-1} \left( \frac{\mu}{ a_\infty^2 r^{D-3}} \right)^\frac{(D-1)(\gamma -1)}{2(D-3)} .
  \label{temp}
\end{equation}
For $D=4$ this reduces to
\begin{equation}
\frac{T(r)}{T_{\infty}} = \left[ \frac{1}{\sqrt{2}} \lambda_s \left(\frac{GM}{a_\infty^2 r}\right)^\frac32 \right]^{\gamma -1}
 \label{temp4d},
\end{equation}
where Shapiro and Teukolsky's accretion eigenvalue $\lambda_s$ is related to our parameter via $\lambda_s = 4\lambda \mid_{D=4}$.

\subsection{Event horizon}
At the event horizon $r = r_H = \mu^\frac{1}{D-3}$. Since the flow is supersonic as we are well below the Bondi radius, the fluid velocity
is still well approximated  by $v^2 \approx \frac{\mu}{r^{D-3}}$. At $r_H$, $v_{H}^{2} \equiv v^2(r_H) \approx 1 $, i.e. the flow speed at the horizon equals the speed of light. Using (\ref{bondi}), (\ref{mdotexplicit}) and (\ref{gasvel}) we obtain the gas compression at the event horizon:
 \begin{equation}
 \frac{n_H}{n_{\infty}} = \lambda \left( \frac{c}{a_\infty} \right)^\frac{D-1}{D-3} , \label{eventcomp}
\end{equation}
where we have re-introduced the speed of light $c$.
Again assuming a Maxwell-Boltzmann gas, $p = n k_B T$, we find the adiabatic temperature profile at the event horizon using (\ref{eos}) and (\ref{eventcomp}): \begin{equation}
\frac{T_H}{T_{\infty}} = \left[ \lambda \left(\frac{c}{a_{\infty}} \right)^\frac{D-1}{D-3}\right]^{\gamma -1}. \label{temphorizon}
\end{equation}
Again in the limit $D=4$, Equation (\ref{temphorizon}) takes the form
\begin{equation}
\frac{T_H}{T_{\infty}} \equiv  \left[\frac{\lambda}{4} \left(\frac{c}{a_{\infty}} \right)^3  \right]^{\gamma-1} \label{temphorizon4d}
\end{equation}
after reinserting the speed of light ($c$) in the above expressions.

\begin{widetext}

\begin{figure}
\begin{tabular}{|c|c|c|}
\hline
\includegraphics[scale=0.5]{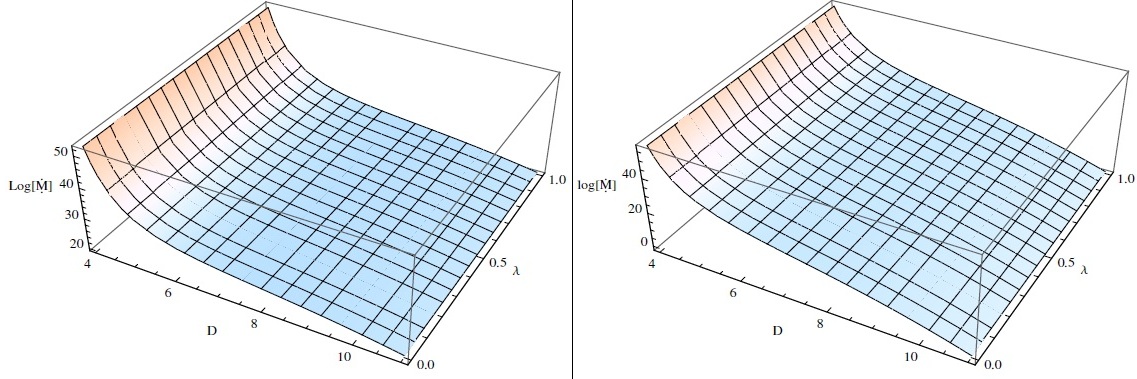} 
\\
\hline
\end{tabular}
\caption{\label{AH} Plots showing the accretion rate $\dot{M}$ as a function of $\lambda_s$ and $D$ for $\gamma=1.1$ (left) and  $\gamma=4/3$ (right). }
\end{figure}

\begin{figure}
\begin{tabular}{|c|c|c|}
\hline
\includegraphics[scale=0.5]{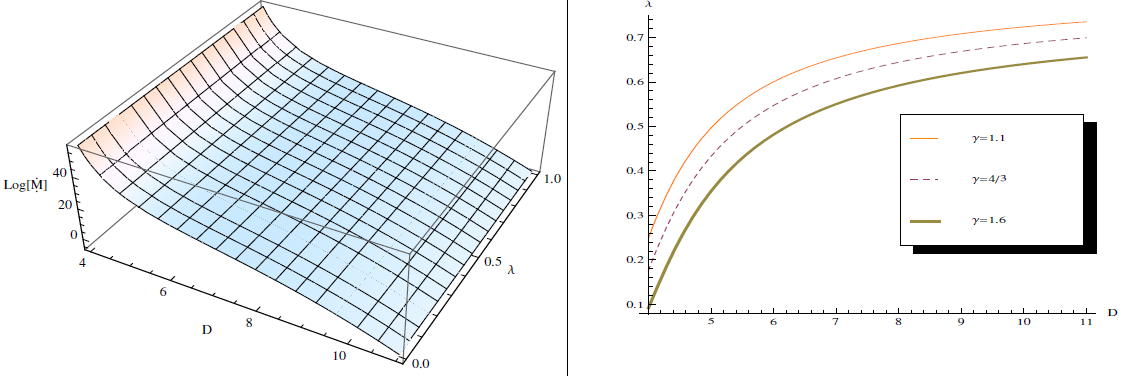} 
\\
\hline
\end{tabular}
\caption{\label{AMD3EV} Left plot shows the accretion rate $\dot{M}$ as a function of $\lambda_s$ and $D$ for $\gamma = 1.6$; Right plot shows the accretion parameter $\lambda_s $ as a function of $D$.}
\end{figure}

\end{widetext}

\section{Conclusions}
In recent years, black hole solutions in more  than four spacetime
dimensions have been the subject of intensive research, motivated by
ideas in brane-world cosmology, string theory and gauge/gravity
duality. Several interesting and surprising results have been found
\cite{horwitz}. In dimensions higher than four, the uniqueness
theorems do not hold due to the fact that there are more degrees of
freedom.  The discovery of black-ring solutions in five dimensions
shows that non-trivial topologies are allowed in higher dimensions
\cite{empran}.  To determine the fate of black holes in higher
dimensional scenarios we considered spherically symmetric, steady
state, adiabatic accretion onto a higher dimensional Schwarzschild
black hole.

We determined the general analytic expressions for the critical
radius and mass accretion rate, for polytropic matter accreting onto
a  $D$-dimensional Schwarzschild black hole. We also found explict
expressions for the gas compression and temperature profile both
below the critical radius and at the event horizon.  The accretion
rate $\dot{M}$ is clearly dependent on the mass and dimensionality
of the black hole. This is to be contrasted with the result of Bondi
\cite{bondi} which showed that $\dot{M} \sim M^2$. Our result also
generalises the study of Giddings and Mangano \cite{tev} which
obtained the mass-dependent accretion rate of matter accreting via
the Newtonian gravity potential of a $D$-dimensional TeV black hole.

We have not considered compactification  of higher dimensions and
leave this as a future project. Upper bounds for higher dimensions
have been established in the literature and their effects on black
hole accretion, as well as other physical processes, will be
restricted to the compactification scale. Beyond this length scale
we expect conventional $4$-dimensional physics to dominate.

A number of extensions to our study of higher dimensional  accretion
are possible. One can attempt to work out the effect of extra
dimensions on the luminosity, frequency spectrum and energy
conversion efficiency of the the accretion flow. More exotic matter,
like scalar fields, could be investigated. Unlike general
relativity, Lovelock gravity and its special case,
Einstein-Gauss-Bonnet gravity, have been demonstrated to be low
energy limits of particular string theories. It may be feasible to
study the effects of accretion on to higher dimensional black holes
described by those gravity theories.

\section{Acknowledgements}
AJJ thanks the NRF and UKZN for financial support. SGG  thanks the University
Grant Commission (UGC) major research project grant F. NO.
39-459/2010 (SR). SDM acknowledges that this work is based upon
research supported by the South African Research Chair Initiative of
the Department of Science and Technology and the National Research
Foundation.


\begin{thebibliography}{99}




\bibitem{bondi}
H. Bondi, Mon. Not. R. Astron. Soc. {\bf 112}, 195 (1952).




\bibitem{bhoyle}
H. Bondi and F. Hoyle, Mon. Not. R. Astron. Soc. {\bf 104}, 273 (1944).




\bibitem{lyttleton}
F. Hoyle and R. A. Lyttleton, Proc. Camb. Phil. Soc. {\bf 35}, 405 (1939).




 \bibitem{edgar}
R. Edgar, New Astronomy Reviews {\bf 48}, 843 (2004).




\bibitem{frank}
J. Frank, A. King and D. Raine, \textit{Accretion power in astrophysics}, 3rd edition (Cambridge University Press, Cambridge, 2002).




\bibitem{st}
S. L. Shapiro and S. A. Teukolsky, \textit{Black Holes, White Dwarfs and Neutron Stars} (Wiley, New York, 1983).




\bibitem{shu}
F. H. Shu, \textit{The Physics of Astrophysics: Gas Dynamics}, Volume II (University Science Books, California, 1992).








\bibitem{michel}
F. C. Michel, Astrophys. Space Sci. {\bf 15}, 153 (1972).




\bibitem{begelman}  M. C. Begelman, Astron. Astrophys. {\bf 70}, 583 (1978).




\bibitem{shap73a}
S. L. Shapiro, Astrophys. J. {\bf 180}, 531 (1973).




\bibitem{shap73b}
S. L. Shapiro, Astrophys. J. {\bf 185}, 69 (1973).




\bibitem{shap74}
S. L. Shapiro, Astrophys. J. {\bf 189}, 343 (1974).




\bibitem{blum} G. R. Blumenthal and W. G. Mathews, Astrophys. J. {\bf 203}, 714 (1976).




\bibitem{brink} W. Brinkmann, Astron.  Astrophys., {\bf 85}, 146 (1980).




\bibitem{malec} E. Malec, Phys. Rev. D {\bf 60}, 104043 (1999).




\bibitem{charge} J. A. de Freitas Pacheco, Journal of Thermodynamics {\bf 2012}, 791870 (2012).




\bibitem{mp} R. C. Myers and M. J. Perry, Annals Phys. {\bf 172}, 304 (1986)




\bibitem{horwitz}
Gary T. Horowitz, \textit{Black Holes in Higher Dimensions}, (Cambridge University Press, Cambridge, England, 2012).




\bibitem{empran} R. Emparan and S. R. Harvey {\it Black Holes in Higher Dimensions, Living Rev. Relativity} {\bf 11},  (2008).




\bibitem{tev}
S. B. Giddings and M. L. Mangano, Phys. Rev. D {\bf 78}, 035009 (2008).




\bibitem{phantom}
M. Sharif and G. Abbas, Mod. Phys. Lett. A {\bf 26}, 1731 (2011).




\bibitem{babprl}
E. Babichev, V. Dokuchaev and Y. Eroshenko, Phys. Rev. Lett. {\bf 93}, 021102 (2004).




\bibitem{phantomrn}
E. Babichev, V. Dokuchaev and Y. Eroshenko, J. Exp. Theor. Phys. {\bf 112}, 784 (2005).








\bibitem{jamhus}
M. Jamil and I. Hussain, Int. J. Theor. Phys. {\bf 50}, 465 (2011).




\bibitem{btz}
M. Jamil and M. Akbar, Gen. Relat. Gravit. {\bf 43}, 1061 (2011).












\bibitem{tangherlini1963schwarzschild}
F. R. Tangherlini, Nuovo Cimento {\bf 27}, 636 (1963).








\end{thebibliography}
\end{document}